\documentclass[a4,amssymb,amsmath,aps,prl,twocolumn,superscriptaddress,showpacs]{revtex4}
\usepackage{graphicx}
\usepackage{dcolumn}
\usepackage{bm}
\usepackage{units}
\usepackage{color}

\begin{document}
\preprint{Lu2V2O7}
\title{Spin wave spectrum of the quantum ferromagnet on the pyrochlore lattice Lu$_2$V$_2$O$_7$}

\author{M. Mena}
\affiliation{London Centre for Nanotechnology and Department of Physics and Astronomy, University College London, Gower Street, London WC1E 6BT, United Kingdom}
\affiliation{Laboratory for Neutron Scattering and Imaging, Paul Scherrer Institute, CH-5232 Villigen PSI, Switzerland}
\author{R. S. Perry}
\affiliation{London Centre for Nanotechnology and Department of Physics and Astronomy, University College London, Gower Street, London WC1E 6BT, United Kingdom}
\affiliation{Centre for Science at Extreme Conditions, University of Edinburgh, Mayfield Road, Edinburgh EH9 3JZ, Scotland}
\author{T. G. Perring}
\affiliation{ISIS Facility, Rutherford Appleton Laboratory, Chilton, Didcot, Oxfordshire OX11 0QX, United Kingdom}
\author{M. D. Le}
\affiliation{Helmholtz-Zentrum Berlin, Hahn-Meitner-Platz 1, D-14109, Berlin, Germany}
\author{S. Guerrero}
\affiliation{Condensed Matter Theory, Paul Scherrer Institute, CH-5232 Villigen, Switzerland}
\author{M. Storni}
\affiliation{Condensed Matter Theory, Paul Scherrer Institute, CH-5232 Villigen, Switzerland}
\author{D. T. Adroja}
\affiliation{ISIS Facility, Rutherford Appleton Laboratory, Chilton, Didcot, Oxfordshire OX11 0QX, United Kingdom}
\author{Ch. R\"uegg}
\affiliation{Laboratory for Neutron Scattering, Paul Scherrer Institute, CH-5232 Villigen PSI, Switzerland}
\affiliation{DPMC-MaNEP, University of Geneva, CH-1211 Geneva 4, Switzerland}
\author{D. F. McMorrow}
\affiliation{London Centre for Nanotechnology and Department of Physics and Astronomy, University College London, Gower Street, London WC1E 6BT, United Kingdom}

\date{\today}

\begin{abstract}
Neutron inelastic scattering has been used to probe the spin dynamics of the quantum ($S$=1/2)
ferromagnet on the pyrochlore lattice Lu$_2$V$_2$O$_7$. Well-defined spin waves are observed at all energies and 
wavevectors, allowing us to determine the parameters of the Hamiltonian of the system. The data are found to be in excellent overall agreement 
with a minimal model that includes a nearest-neighbour Heisenberg exchange $J\!=\!8.22(2)$~meV and a Dzyaloshinskii-Moriya interaction (DMI) $D\!=\!1.5(1)$~meV. 
The large DMI term revealed by our study is broadly consistent with the model developed by 
Onose {\it et al.}  to explain  the magnon Hall effect they observed
in  Lu$_2$V$_2$O$_7$ \cite{onose2010}, although our ratio of $D/J=0.18(1)$ is roughly half of their value and three times larger than calculated by \textit{ab initio} methods \cite{xiang2011}. 
\end{abstract}

\maketitle
\bigskip

The highly frustrated nature of the pyrochlore lattice leads to a rich diversity of fascinating properties when the lattice sites are decorated with ``classical'' (large $S$) spins \cite{gardner2010}. Arguably the most celebrated example is ferromagnetically coupled Ising spins which give rise to spin-ice \cite{harris1997,bramwell2001} and the emergence of magnetic monopoles \cite{castelnovo2008,bramwell2009}. While many examples of classical pyrochlores exist, there are few examples of pyrochlores where the spins of the magnetic ions are explicitly in the quantum ($S$=1/2) limit. Quantum effects can, however, play a decisive role even in classical pyrochlores if their low-energy physics maps onto an effective spin 1/2 model \cite{gardner2010}. In either case,  quantum effects may produce a range of novel phenomena including the realisation of  a three-dimensional quantum spin-liquid ground state, emergent electromagnetism supporting photon-like excitations, {\it etc.} \cite{harris1991,hermele2004,ross2011,benton2012}. Interest in itinerant pyrochlore magnets is also 
motivated by the various anomalous transport properties they exhibit \cite{nagaosa2010}. 

Lu$_2$V$_2$O$_7$ is a ferromagnetic, small-gap Mott insulator that crystallises in the pyrochlore structure and 
displays a number of exceptional properties. Fig. 1(a) shows the V$^{4+}$ ($S$=1/2) sites in the pyrochlore lattice, which form a three-dimensionally coordinated network of corner sharing tetrahedra.
Bulk measurements have established that the spins order ferromagnetically at $T_C$=70 K \cite{onose2010,zhou2008}.
Measurements of the thermal conductivity in Lu$_2$V$_2$O$_7$  by Onose {\it et al.} have been interpreted in terms of \emph{a magnon Hall effect} \cite{onose2010}, based on the observation that the thermal conductivity has a distinctive dependence on applied magnetic fields for temperatures below $T_C$. This highly unusual and previously unreported phenomenon was shown to be consistent with a model in which the Dzyaloshinskii-Moriya interaction (DMI) between nearest neighbour spins deflects magnon wavepackets propagating from the hot to the cold side of the sample \cite{onose2010}. 

Further evidence of the exceptional properties of Lu$_2$V$_2$O$_7$ was provided by Zhou et al. \cite{zhou2008} who discovered that it displays a large (50\%) magneto-resistance around 70 K in an applied magnetic field. Additionally, polarised neutron diffraction found evidence for an orbital ordered ground state with associated super-exchange pathways favouring a dominant nearest-neighbour ferromagnetic coupling \cite{ichikawa2005}.  
It  has also been suggested that Lu$_2$V$_2$O$_7$ may be an example of a topological magnon insulator with chiral 
edge states \cite{zhang2013}. 

\begin{figure}
	\includegraphics[width=8.78cm]{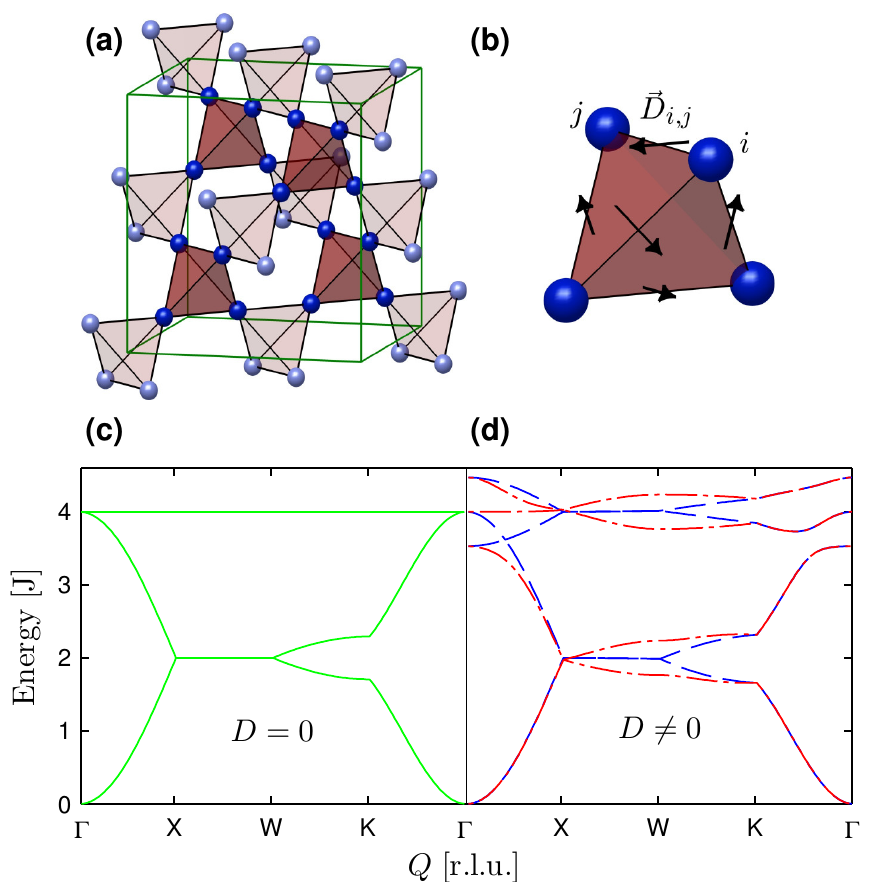}
	\caption{\label{fig_struct}(a) Pyrochlore structure of Lu$_2$V$_2$O$_7$. The 
	 V$^{4+}$ ions (blue) form a network of vertex-sharing tetrahedra with four tetrahedra in the unit cell. (b) A single tetrahedron. The arrows located between adjacent V$^{4+}$ sites 
	 denote the directions of the DMI vectors, e.g. the vector $\vec{D}_{i,j}$ is associated with the interaction between sites $i$ and $j$. 
	 Calculated spin-wave dispersion for Lu$_2$V$_2$O$_7$ for (c) $D$ = 0 and (d) $D = J/3$ with the local magnetic field $\vec{H}$ parallel to either (100) (blue) or (010) (red). $\Gamma$, X, W and K label the 
	 high-symmetry points of the conventional FCC Brillouin zone.}
\end{figure}

The magnon Hall effect data in Lu$_2$V$_2$O$_7$ has been interpreted in terms of a Hamiltonian that takes the form
\begin{equation} \label{H_Luvo1}
\mathcal{H} =  \sum_{<ij>} \big( - J \vec{S_i} \cdot \vec{S_j} + \vec{D}_{ij} \cdot (\vec{S_i} \times \vec{S_j}) \big)+ g\mu_B    \sum_i \vec{H}\cdot \vec{S_i} 
\end{equation}
where  $J$ is the nearest-neighbour Heisenberg exchange, $\vec{D}_{ij}$ represents the DMI, whose directions are sketched in Fig. 1(b), 
$\vec{S}_{i}$ are spin operators, $\vec{H}$ is the magnetic field, and $<$\textit{ij}$>$ runs over all pairs of nearest-neighbours \cite{onose2010}.
To date there have been no reports of experiments designed to determine accurately the parameters of the 
Hamiltonian. 
Following standard procedures, this Hamiltonian can be diagonalized by expressing the spin operators in the local coordinates of each of the four ferromagnetic sublattices and applying the Holstein-Primakoff transformation. Figure \ref{fig_struct}(c)-(d) show the calculated dispersion for the four branches for different values of $D=|\vec{D}_{ij}|$ 
and $\vec{H}$.

The magnitude of the anomalous contribution to the thermal conductivity due to magnons is determined by the 
ratio  $D/J$, which was estimated  in Lu$_2$V$_2$O$_7$ to be $D/J\!\simeq\!1/3$ by fitting the transport data \cite{onose2010} and hence would require strong spin-orbit coupling. 
In contrast, DFT calculations obtained a much lower ratio $D/J\!\simeq\!1/20$, with the authors instead emphasising the importance of single ion anisotropy \cite{xiang2011}. There is thus a  {\it prima facie} case for accurately determining the
Hamiltonian of Lu$_2$V$_2$O$_7$ both in terms of interpreting the transport data, and more generally of 
understanding the spin-dynamics of this elusive example of a pyrochlore lattice in the quantum limit.
Here we present the results of a neutron inelastic scattering experiment that satisfies these objectives.

Lu$_2$V$_2$O$_7$ crystallises in the cubic \textit{Fd$\bar{3}$m} space group (number 227), with a 
lattice parameter of 9.94 \AA.
Single crystals were grown in an image furnace, and were characterised by SQUID magnetometry and X-ray diffraction, confirming the ferromagnetic transition temperature and good crystalline quality. 
Two single crystals, total mass of 3.6 g, were co-aligned with the (\textit{HHL}) plane horizontal. 
The experiment was performed on the MERLIN direct geometry, time-of-flight spectrometer at the ISIS facility (UK)\cite{bewley2006}.  Data were collected for incoming energies $E_i$ (measured elastic resolutions, FWHM) of 25 (3.0), 50 (5.3) and 80 (7.2) meV and at a temperature of 4 K. 
The small moment size and mass of the sample necessitated typical counting times of 15 hours to acquire sufficient statistics 
for a given energy and angular setting of the sample.
Data sets were collected with the incident neutron beam along the $<$001$>$, $<$110$>$ and $<$111$>$ directions. The data were corrected for detector efficiency, outcoming versus incoming wavevector ratio $k_f/k_i$ and normalised to a vanadium standard using the program MANTID \cite{mantid}. The resulting S($\vec Q$,w) data sets were analysed with the HORACE software package \cite{horace}.

\begin{figure}[t!]
	\includegraphics[width=8.78cm]{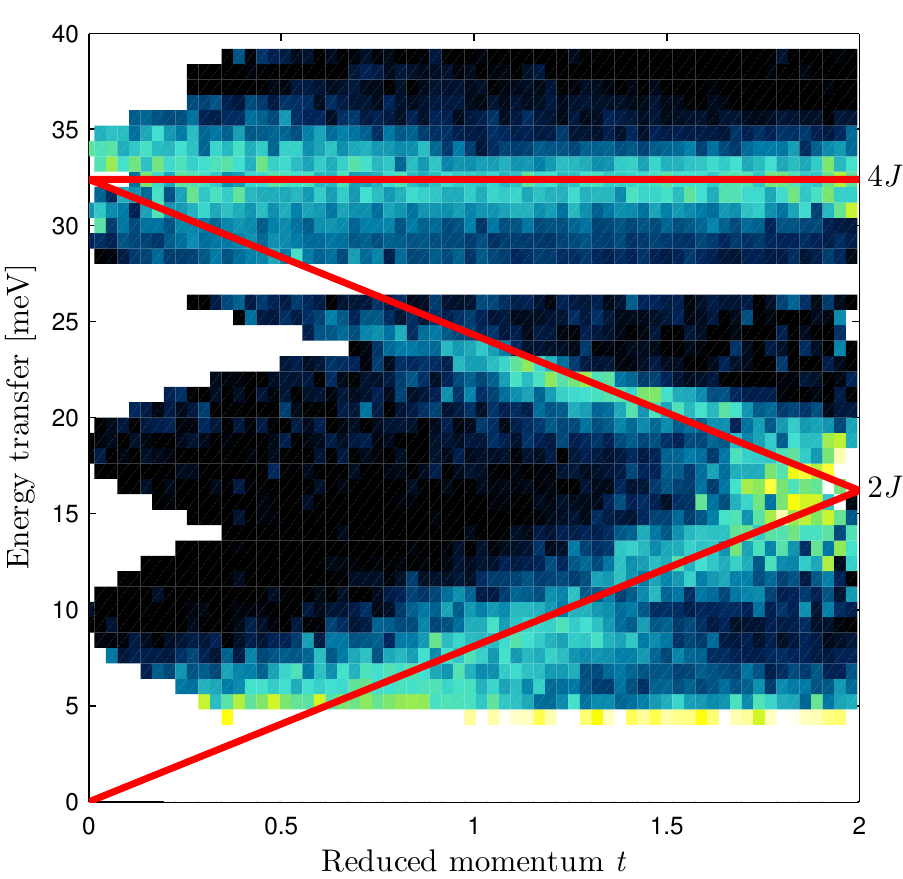}
	\caption{\label{fig_reduced} Neutron inelastic scattering data ($E_i = 50$ meV) from Lu$_2$V$_2$O$_7$ plotted as 
	a function of the reduced momentum $t$ defined in the text. The data between 0 and 27 meV energy transfer (below the white gap) are obtained by averaging over data with $|\vec{Q}| <$ 3 r.l.u., while between 28 and 40 meV energy transfer (above the white gap) they are obtained by considering only data with $|\vec{Q}| <$ 5 r.l.u. The solid red line represents a fit of the dispersion with $J$ = 8.1(1) meV and $D=0$ meV.}
\end{figure}

In Fig.\ \ref{fig_reduced} we show an overview of the data covering the full energy range over which we observed magnetic 
scattering. For the abscissa we have chosen to use a reduced wavevector coordinate, \textit{t}. This allows us
to exploit the symmetry of the system, and to utilise data from more than one Brillouin zone and orientation of the crystal. The reduced 
coordinate $t$ is defined by $t(\vec{Q})=2-(1+\cos{(\pi H)}\cos{(\pi K)}+\cos{(\pi K)}\cos{(\pi L)}+\cos{(\pi L)}\cos{(\pi H)}) ^{1/2}$ where $t \in [0,2]$. 
Re-binning the data in terms of $t$ folds several Brillouin zones onto the same axis; for example, any zone centre will be re-binned to $t=0$.  In order to exclude scattering from aluminium phonons from the sample environment only data with $|\vec{Q}|<3$ r.l.u. were averaged below 27 meV, and $|\vec{Q}|<5$ r.l.u. between 27 and 40 meV.
For $D=0$ meV the dispersions of the four modes associated with the four ferromagnetic sublattices of
a tetrahedron become $\hbar \omega_{1}(t)= t J$,  $\hbar \omega_{2}(t)=(4-t) J$, $\hbar\omega_{3,4}(t)=4J$ \cite{onose2010}. The solid line in Fig.\ \ref{fig_reduced} represents a fit of this dispersion to the data from
which we obtained the nearest-neighbour exchange coupling $J$ = 8.1(1) meV. Thus even at this 
level of analysis we can assert that the Hamiltonian relevant to Lu$_2$V$_2$O$_7$ is dominated by 
isotropic nearest-neighbour exchange. However, the data analysis performed in this way, while useful for providing an overview, does not lend itself to a full quantitative treatment since sampling is performed over many Brillouin zones and can lead to distortions of the data both from extrinsic (phonon contamination, spurious scattering, etc.) and resolution effects.

\begin{figure}
	\includegraphics[width=8.0cm]{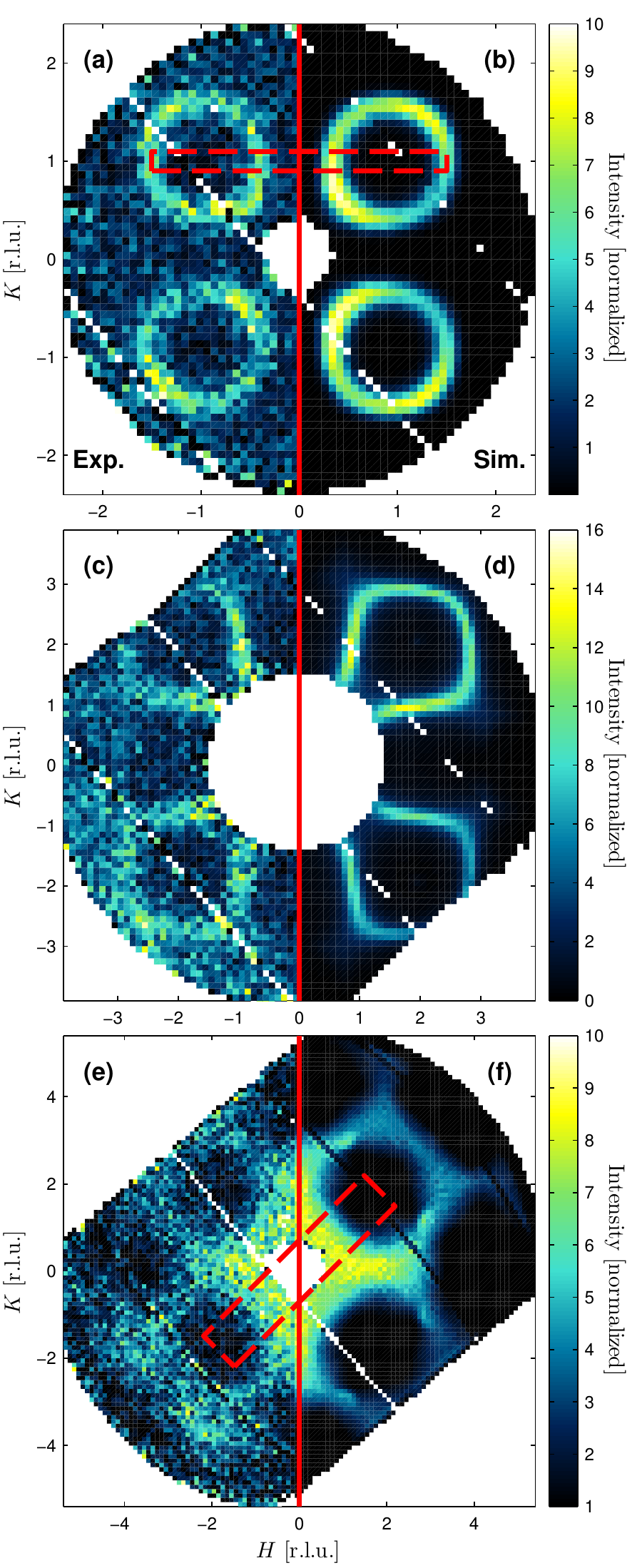}
	\caption{\label{fig_datsim} Comparison of data (left half) and MCPHASE simulations (right half). (a) $E_i$ = 25 meV, averaged with $L$ varying between 0.9 and 1.1 r.l.u. (c) $E_i$ = 50 meV, averaged between 1.9-2.1 r.l.u. in $L$. (e) $E_i$ = 80 meV, averaged between 31-35 meV in energy. A background has been subtracted for this data set only. Energy varies from 12meV at the center to 4meV at the edges in (a)-(b), 20meV to 10meV in (c)-(d). L varies from 4.5 in the center to 2 at the edges in (e)-(f). (b), (d) and (f) are simulations performed with $J$ = 8.1 meV and $D$ = 0 meV. Dashed boxes represent regions of interest analysed in cuts shown in  Fig.\ \ref{fig_fits}.}
\end{figure}

\begin{figure}
\includegraphics[width=8.78cm]{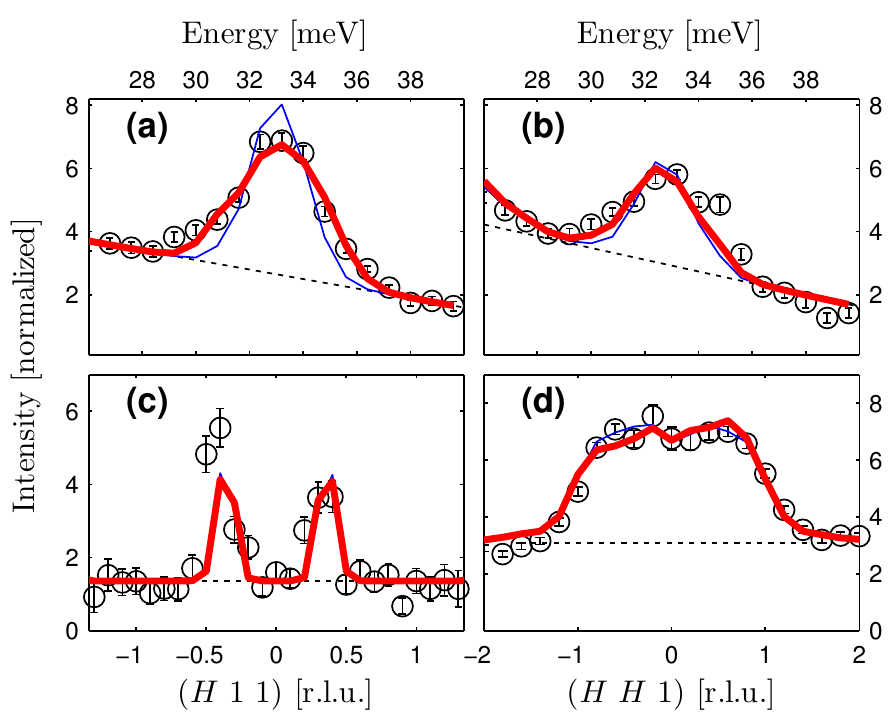}
\caption{\label{fig_fits} 
Analysis of specific regions of $(\vec{Q},\omega)$ to determine the best values of $J$  and $D$.
Red lines represent the best fit of the MCPHASE simulations with $J = 8.22(2)$ meV and $D = 1.5(1)$ meV;  blue lines, 
$J = 8.22(2)$ meV and $D = 0$ meV; black lines, background. (a) and (b): Energy scans at $\sim4J$  obtained by averaging data over 
$1.5 < t(\vec{Q}) < 1.7$ and $0.7 < t(\vec{Q}) < 0.9$, respectively. (c)  Wavevector dependence
along H averaged for $0.9\leq K \leq 1.1$, $0.9\leq L \leq 1.1$ and $6\leq E \leq 8$ (box, Fig.\ \ \ref{fig_datsim}(a)-(b)).
(d) Wavevector dependence along $(H H 1)$ averaged for $-0.5\leq (K \bar{K} 1) \leq 0.5$, $0\leq L \leq 6$ and $28\leq E \leq 35$ (box, Fig.\ \ \ref{fig_datsim}(e)-(f)). }
\end{figure}

Further data analysis and modelling were therefore performed in natural reciprocal space coordinates $(H,K,L)$.
Figure \ref{fig_datsim} shows a compendium of data (left hand panels, (a), (c) and (e)) plotted as a function of $H$ and $K$ for different intervals of both energy and $L$.
For energies lower than approximately 16 meV$\approx 2J$ (Fig.\ \ref{fig_datsim}(a)), rings of intensity are evident, centred on (1 1 1)-type $\Gamma$-points, as expected from the almost-quadratic form of the dispersion at low energies, see Figs.\ \ref{fig_struct}(c) and (d).
At energies $\sim 2J$ (Fig.\ \ref{fig_datsim}(c)), the signal has moved to the edges of the Brillouin zones. 
For energies $\sim 4J$ (Fig.\ \ref{fig_datsim}(e)) the dispersion is expected to be almost flat -- with maximal deviations $\pm \sqrt{2}D$ -- and the significant $\vec{Q}$-dependence of the spectral weight gives rise to a square lattice-like structure in the intensity.
(In Fig.\ \ref{fig_datsim}(e) a smoothly varying $|\vec{Q}|$-dependent background was subtracted from the data.) 

Simulations of the data were performed using the package MCPHASE \cite{mcphase}, which uses a mean-field random-phase approximation to calculate the energies and spectral weights of the magnetic excitations, which were then convoluted with a Gaussian of the same width as the calculated instrumental energy resolution at the appropriate energy transfer in order to model $S(\vec{Q},\omega)$. MCPHASE was configured to simulate $S(\vec{Q},\omega)$ expected from Eq.\ 1, with an isotropic spin-only magnetic form factor for the V$^{4+}$ ions calculated within the dipole approximation.
In Figs.\ \ref{fig_datsim}(b), (d) and (f) simulations are shown for $J$ = 8.1 meV and $D$ = 0 meV.
Remarkably, this minimal model which neglects the DMI and uses an isotropic form factor provides an excellent description of the data over all relevant wavevectors and energies.
It  is therefore apparent that to a good approximation the Hamiltonian in Lu$_2$V$_2$O$_7$ is dominated by nearest-neighbour Heisenberg exchange. Simulations of  $S(\vec{Q},\omega)$ were also performed using the form factor for
the putative orbital  ordered state \cite{ichikawa2005}: over the measured range of $(\vec{Q},\omega)$ the results were essentially identical to those performed using the isotropic form factor. 

However, it is apparent that the effects of the DMI are more pronounced in specific regions of the spin-wave dispersion, 
Fig.\ \ref{fig_struct}, and may not necessarily reveal themselves in cuts of the type shown in Fig.\ \ref{fig_datsim}.
We thus analysed data from specific regions of the spin-wave dispersion where the effects of the 
DMI were expected to be more significant.
Figure \ref{fig_fits} shows data extracted as cuts in both energy ((a) and (b)) and wavevector ((c) and (d)).
MCPHASE simulations were then fitted to this data to explore the dependence of the calculated scattering on $J$ and $D$.
One complication in performing simulations for finite values of $D$ is that, 
as shown in Fig.\ \ref{fig_struct}(d),  the spin-wave energy is dependent on the field direction. 
As no external field was applied in the experiment, in the simulations we assumed domain coexistence and 
averaged the simulations over a set of easy-axis $<$100$>$-type domains. 

For our data set, greatest sensitivity to the values of $J$ and $D$, with minimal contamination from phonon scattering, 
was obtained in energy cuts for energies $\sim 4J$. The best fits of the MCPHASE simulations to energy scans 
around $\sim 4J$, represented by the red-lines in Fig.\ \ref{fig_fits}(a) and (b), were obtained with $J = 8.22(2)$ meV and $D = 1.5(1)$ meV, with 
$\chi^2$=2.9 and $\chi^2$=6.7 for (a) and (b), respectively. Constraining the value of $D=0$ produced significantly worse fits, as represented by the blue lines, with  $\chi^2$=14 and $\chi^2$=13 for (a) and (b), respectively. The optimal values were found to be consistent with data analysed from other regions of $(\vec{Q},\omega)$, albeit with less sensitivity to $D$, as illustrated in  Fig.\ \ref{fig_fits}(c) and (d).

The ratio of  $D/J=0.18(1)$ determined in our study establishes the existence of a large finite DMI in Lu$_2$V$_2$O$_7$.
Its value is roughly half the ratio of $\sim$1/3 used in the analysis of the magnon Hall effect data \cite{onose2010}. 
It is clearly substantially larger than the value 0.05 obtained from DFT calculations \cite{xiang2011}.
However, our determination of the leading term in the Hamiltonian, $J$=8.22(2) meV, is  in reasonable accord with the corresponding value of   
$J$=7.1 meV from DFT taking into account the accuracy of the computational method.

In conclusion, we have performed a neutron inelastic scattering experiment on the quantum ferromagnetic pyrochlore 
Lu$_2$V$_2$O$_7$. The well-defined  spin-wave dispersion observed at all energies and wavevectors 
has allowed us to determine accurately for the first time the parameters of the
Hamiltonian. Our data is to a remarkable extent accounted for by a minimal
model of isotropic nearest-neighbour Heisenberg exchange with $J\!=\!8.22(2)$ meV and with a DMI given by $D=1.5(1)$ meV. The existence of a large DMI term in Lu$_2$V$_2$O$_7$ may be taken as evidence in favour of the reported interpretation of the magnon Hall effect data\cite{onose2010}. Our value of $D/J$ is roughly half of the value required to explain the thermal transport data within the current model. We hope that our results will stimulate further theoretical studies to explore this specific aspect, as well as wider questions such as the role of the topological band structure, of this  
intriguing material. Our data show that even in the quantum limit, 
well defined spin-waves can propagate on the highly-frustrated pyrochlore lattice in the presence of 
nearest-neighbour ferromagnetic exchange. This situation contrasts starkly with the classical spin ices on the same 
lattice, where the long-range dipolar interaction gives rise to an effective ferromagnetic coupling and monopolar
elementary excitations \cite{castelnovo2008,bramwell2009}. 

The research was supported by the EPSRC and the Paul Scherrer Institute, and part of the research leading to these results
has received funding from the European Community's Seventh Framework Programme (FP7/2007-2013) under grant agreement no 312284.
We thank Steve Bramwell for useful discussions, and acknowledge the help of Zhuo Feng, Simon Ward and the ISIS technicians.


\begin{thebibliography}{9}

\bibitem{onose2010}
Y. Onose, T. Ideue, H. Katsura, Y. Shiomi, N. Nagaosa, Y. Tokura, Science \textbf{329}, 297 (2010).

\bibitem{xiang2011}
H. J. Xiang, E. J. Kan, M.-H. Whangbo, C. Lee, Su-Huai Wei, X. G. Gong, Pys. Rev. B \textbf{83}, 174402 (2011).

\bibitem{gardner2010}
J.S. Gardner, M.J.P. Gingras, J.E. Greedan, Rev. Mod. Phys., \textbf{82}, 53 (2010).

\bibitem{harris1997}
M.J. Harris, S.T. Bramwell, D.F. McMorrow, T. Zeiske, K.W. Godfrey, Phys. Rev. Lett. \textbf{79}, 2554 (1997).

\bibitem{bramwell2001}
S.T. Bramwell, M.J.P. Gingras, Science \textbf{294}, 5546 (2001).

\bibitem{castelnovo2008}
C. Castelnovo, R. Moessner, and S.L. Sondhi, Nature \textbf{451} ,42 (2008).

\bibitem{bramwell2009}
S.T. Bramwell, S.R. Giblin, S. Calder, R. Aldus, D. Prabhakaran, and T. Fennell, Nature \textbf{461}, 956 (2009).

\bibitem{harris1991}
A.B. Harris,  A.J. Berlinsky, and C. Bruder,  J. Appl. Phys., \textbf{69}, 5200. (1991).

\bibitem{hermele2004}
M. Hermele, M.P.A. Fisher, and L. Balents, Phys. Rev. B \textbf{69}, 064404 (2004).

\bibitem{ross2011}
K.A. Ross, L. Savary, B.D. Gaulin, and L. Balents, Phys. Rev. X 1, 021002 (2011).
 
\bibitem{benton2012}
O. Benton, O. Sikora, N. Shannon, Phys. Rev. B. \textbf{86}, 075154 (2012).

\bibitem{nagaosa2010}
N. Nagaosa, J. Sinova, S. Onoda, A.H. McDonald, and N.P. Ong,  Rev. Modern Phys., \textbf{82} 1539 (2010).

\bibitem{zhou2008}
H. D. Zhou, E. S. Choi, J. A. Souza, J. Lu, Y. Xin, L. L. Lumata, B. S. Conner, L. Balicas, J. S. Brooks, J. J. Neumeier, and C. R. Wiebe; Phys. Rev. B \textbf{77}, 020411 (2008).

\bibitem{ichikawa2005}
H. Ichikawa, L. Kano, M. Saitoh, S. Miyahara, N. Furukawa, J. Akimatsu, T. Yokoo, T. Matsumura, M. Takeda, K. Hirota, 
J. Phys. Soc. Jap. \textbf{74}, 1020 (2005).

\bibitem{zhang2013}
L. Zhang, J. Ren, J.-S. Wang, B. Li, Phys. Rev. B \textbf{87}, 144101 (2013).

\bibitem{bewley2006}
R. I. Bewley, R. S. Eccleston, K. A. McEwen, S. M. Hayden, M. T. Dove, S. M. Bennington, J .R. Treadgold, R. L. S. Coleman, Physica B \textbf{385}, 1029-1031 (2006).

\bibitem{mantid}
Available at http://www.mantidproject.org.

\bibitem{horace}
Available at http://horace.isis.rl.ac.uk.

\bibitem{mcphase}
M. Rotter, S. Kramp, M. Loewenhaupt, E. Gratz, W. Schmidt, N. M. Pyka, B. Hennion, R.v.d. Kamp; Appl. Phys. A \textbf{74}, 751 (2002).

\end{thebibliography}
\end{document}